\documentclass[a4paper]{llncs}
\usepackage{algpseudocode}
\usepackage{alltt}
\usepackage{amsfonts}
\usepackage[british]{babel}
\usepackage{hyperref}
\usepackage[utf8]{inputenc}
\usepackage{listings}

\renewcommand{\verb}{\lstinline}

\begin{document}

\mainmatter
\title{On the correctness of a branch displacement algorithm\thanks{Research supported by the CerCo project, within the Future and Emerging Technologies (FET) programme of the Seventh Framework Programme for Research of the European Commission, under FET-Open grant number 243881}}
\author{Jaap Boender \and Claudio Sacerdoti Coen}
\institute{Dipartimento di Scienze dell'Informazione,\\ Universit\`a degli Studi di Bologna}

\maketitle

\begin{abstract}
The branch displacement problem is a well-known problem in assembler design.
It revolves around the feature, present in several processor families, of
having different instructions, of different sizes, for jumps of different
displacements. The problem, which is provably NP-hard, is then to select the
instructions such that one ends up with the smallest possible program. 

During our research with the CerCo project on formally verifying a C compiler,
we have implemented and proven correct an algorithm for this problem. In this
paper, we discuss the problem, possible solutions, our specific solutions and
the proofs.

\end{abstract}

\section{Introduction}

The problem of branch displacement optimisation, also known as jump encoding, is
a well-known problem in assembler design~\cite{Hyde2006}. It is caused by the
fact that in many architecture sets, the encoding (and therefore size) of some
instructions depends on the distance to their operand (the instruction 'span').
The branch displacement optimisation problem consists of encoding these
span-dependent instructions in such a way that the resulting program is as
small as possible.

This problem is the subject of the present paper. After introducing the problem
in more detail, we will discuss the solutions used by other compilers, present 
the algorithm we use in the CerCo assembler, and discuss its verification,
that is the proofs of termination and correctness using the Matita proof
assistant~\cite{Asperti2007}.
 
The research presented in this paper has been executed within the CerCo project
which aims at formally verifying a C compiler with cost annotations. The
target architecture for this project is the MCS-51, whose instruction set
contains span-dependent instructions. Furthermore, its maximum addressable
memory size is very small (64 Kb), which makes it important to generate
programs that are as small as possible. 

With this optimisation, however, comes increased complexity and hence
increased possibility for error. We must make sure that the branch instructions
are encoded correctly, otherwise the assembled program will behave
unpredictably.

\section{The branch displacement optimisation problem}

In most modern instruction sets that have them, the only span-dependent
instructions are branch instructions. Taking the ubiquitous x86-64 instruction
set as an example, we find that it contains eleven different forms of the
unconditional branch instruction, all with different ranges, instruction sizes
and semantics (only six are valid in 64-bit mode, for example). Some examples
are shown in Figure~\ref{f:x86jumps} (see also~\cite{IntelDev}).

\begin{figure}[h]
\begin{center}
\begin{tabular}{|l|l|l|}
\hline
Instruction & Size (bytes) & Displacement range \\
\hline
Short jump & 2 & -128 to 127 bytes \\
Relative near jump & 5 & $-2^{32}$ to $2^{32}-1$ bytes \\
Absolute near jump & 6 & one segment (64-bit address) \\
Far jump & 8 & entire memory \\
\hline
\end{tabular}
\end{center}
\caption{List of x86 branch instructions}
\label{f:x86jumps}
\end{figure}

The chosen target architecture of the CerCo project is the Intel MCS-51, which
features three types of branch instructions (or jump instructions; the two terms
are used interchangeably), as shown in Figure~\ref{f:mcs51jumps}.

\begin{figure}[h]
\begin{center}
\begin{tabular}{|l|l|l|l|}
\hline
Instruction & Size    & Execution time & Displacement range \\
            & (bytes) & (cycles) & \\
\hline
SJMP (`short jump') & 2 & 2 & -128 to 127 bytes \\
AJMP (`absolute jump') & 2 & 2 & one segment (11-bit address) \\
LJMP (`long jump') & 3 & 3 & entire memory \\
\hline
\end{tabular}
\end{center}
\caption{List of MCS-51 branch instructions}
\label{f:mcs51jumps}
\end{figure}

Conditional branch instructions are only available in short form, which
means that a conditional branch outside the short address range has to be
encoded using three branch instructions (for instructions whose logical
negation is available, it can be done with two branch instructions, but for
some instructions this is not available); the call instruction is
only available in absolute and long forms.

Note that even though the MCS-51 architecture is much less advanced and simpler
than the x86-64 architecture, the basic types of branch instruction
remain the same: a short jump with a limited range, an intra-segment jump and a
jump that can reach the entire available memory.
 
Generally, in code fed to the assembler as input, the only
difference between branch instructions is semantics, not span. This
means that a distinction is made between an unconditional branch and the
several kinds of conditional branch, but not between their short, absolute or
long variants.

The algorithm used by the assembler to encode these branch instructions into
the different machine instructions is known as the {\em branch displacement
algorithm}. The optimisation problem consists of finding as small an encoding as
possible, thus minimising program length and execution time.

This problem is known to be NP-complete~\cite{Robertson1979,Szymanski1978},
which could make finding an optimal solution very time-consuming. 

The canonical solution, as shown by Szymanski~\cite{Szymanski1978} or more
recently by Dickson~\cite{Dickson2008} for the x86 instruction set, is to use a
fixed point algorithm that starts with the shortest possible encoding (all
branch instruction encoded as short jumps, which is likely not a correct
solution) and then iterates over the program to re-encode those branch
instructions whose target is outside their range.

\subsection*{Adding absolute jumps}

In both papers mentioned above, the encoding of a jump is only dependent on the
distance between the jump and its target: below a certain value a short jump
can be used; above this value the jump must be encoded as a long jump.

Here, termination of the smallest fixed point algorithm is easy to prove. All
branch instructions start out encoded as short jumps, which means that the
distance between any branch instruction and its target is as short as possible.
If, in this situation, there is a branch instruction $b$ whose span is not
within the range for a short jump, we can be sure that we can never reach a
situation where the span of $j$ is so small that it can be encoded as a short
jump. This argument continues to hold throughout the subsequent iterations of
the algorithm: short jumps can change into long jumps, but not \emph{vice versa},
as spans only increase. Hence, the algorithm either terminates early when a fixed
point is reached or when all short jumps have been changed into long jumps.

Also, we can be certain that we have reached an optimal solution: a short jump
is only changed into a long jump if it is absolutely necessary.

However, neither of these claims (termination nor optimality) hold when we add
the absolute jump, as with absolute jumps, the encoding of a branch
instruction no longer depends only on the distance between the branch
instruction and its target: in order for an absolute jump to be possible, they
need to be in the same segment (for the MCS-51, this means that the first 5
bytes of their addresses have to be equal). It is therefore entirely possible
for two branch instructions with the same span to be encoded in different ways
(absolute if the branch instruction and its target are in the same segment,
long if this is not the case).

\begin{figure}[ht]
\begin{alltt}
    jmp X
    \vdots
L\(\sb{0}\):
    \vdots
    jmp L\(\sb{0}\)
\end{alltt}
\caption{Example of a program where a long jump becomes absolute}
\label{f:term_example}
\end{figure}

This invalidates our earlier termination argument: a branch instruction, once encoded
as a long jump, can be re-encoded during a later iteration as an absolute jump.
Consider the program shown in Figure~\ref{f:term_example}. At the start of the
first iteration, both the branch to {\tt X} and the branch to $\mathtt{L}_{0}$
are encoded as small jumps. Let us assume that in this case, the placement of
$\mathtt{L}_{0}$ and the branch to it are such that $\mathtt{L}_{0}$ is just
outside the segment that contains this branch. Let us also assume that the
distance between $\mathtt{L}_{0}$ and the branch to it are too large for the
branch instruction to be encoded as a short jump.

All this means that in the second iteration, the branch to $\mathtt{L}_{0}$ will
be encoded as a long jump. If we assume that the branch to {\tt X} is encoded as
a long jump as well, the size of the branch instruction will increase and
$\mathtt{L}_{0}$ will be `propelled' into the same segment as its branch
instruction, because every subsequent instruction will move one byte forward.
Hence, in the third iteration, the branch to $\mathtt{L}_{0}$ can be encoded as
an absolute jump. At first glance, there is nothing that prevents us from
constructing a configuration where two branch instructions interact in such a
way as to iterate indefinitely between long and absolute encodings.

This situation mirrors the explanation by Szymanski~\cite{Szymanski1978} of why
the branch displacement optimisation problem is NP-complete. In this explanation,
a condition for NP-completeness is the fact that programs be allowed to contain
{\em pathological} jumps. These are branch instructions that can normally not be
encoded as a short(er) jump, but gain this property when some other branch
instructions are encoded as a long(er) jump. This is exactly what happens in
Figure~\ref{f:term_example}. By encoding the first branch instruction as a long
jump, another branch instruction switches from long to absolute (which is
shorter).

In addition, our previous optimality argument no longer holds. Consider the program
shown in Figure~\ref{f:opt_example}. Suppose that the distance between
$\mathtt{L}_{0}$ and $\mathtt{L}_{1}$ is such that if {\tt jmp X} is encoded
as a short jump, there is a segment border just after $\mathtt{L}_{1}$. Let
us also assume that the three branches to $\mathtt{L}_{1}$ are all in the same
segment, but far enough away from $\mathtt{L}_{1}$ that they cannot be encoded
as short jumps.

\begin{figure}[ht]
\begin{alltt}
L\(\sb{0}\): jmp X
X:
    \vdots
L\(\sb{1}\):
    \vdots
    jmp L\(\sb{1}\)
    \vdots
    jmp L\(\sb{1}\)
    \vdots
    jmp L\(\sb{1}\) 
    \vdots
\end{alltt}
\caption{Example of a program where the fixed-point algorithm is not optimal}
\label{f:opt_example}
\end{figure}

Then, if {\tt jmp X} were to be encoded as a short jump, which is clearly
possible, all of the branches to $\mathtt{L}_{1}$ would have to be encoded as
long jumps. However, if {\tt jmp X} were to be encoded as a long jump, and
therefore increase in size, $\mathtt{L}_{1}$ would be `propelled' across the
segment border, so that the three branches to $\mathtt{L}_{1}$ could be encoded
as absolute jumps. Depending on the relative sizes of long and absolute jumps,
this solution might actually be smaller than the one reached by the smallest
fixed point algorithm.

\section{Our algorithm}

\subsection{Design decisions}

Given the NP-completeness of the problem, to arrive at an optimal solution
(using, for example, a constraint solver) will potentially take a great amount
of time. 

The SDCC compiler~\cite{SDCC2011}, which has a backend targetting the MCS-51
instruction set, simply encodes every branch instruction as a long jump
without taking the distance into account. While certainly correct (the long
jump can reach any destination in memory) and a very fast solution to compute,
it results in a less than optimal solution. 

On the other hand, the {\tt gcc} compiler suite~\cite{GCC2012}, while compiling
C on the x86 architecture, uses a greatest fix point algorithm. In other words,
it starts with all branch instructions encoded as the largest jumps
available, and then tries to reduce the size of branch instructions as much as
possible.

Such an algorithm has the advantage that any intermediate result it returns
is correct: the solution where every branch instruction is encoded as a large
jump is always possible, and the algorithm only reduces those branch
instructions whose destination address is in range for a shorter jump.
The algorithm can thus be stopped after a determined number of steps without
sacrificing correctness.

The result, however, is not necessarily optimal. Even if the algorithm is run
until it terminates naturally, the fixed point reached is the {\em greatest}
fixed point, not the least fixed point. Furthermore, {\tt gcc} (at least for
the x86 architecture) only uses short and long jumps. This makes the algorithm
more efficient, as shown in the previous section, but also results in a less
optimal solution.

In the CerCo assembler, we opted at first for a least fixed point algorithm,
taking absolute jumps into account.

Here, we ran into a problem with proving termination, as explained in the
previous section: if we only take short and long jumps into account, the jump
encoding can only switch from short to long, but never in the other direction.
When we add absolute jumps, however, it is theoretically possible for a branch
instruction to switch from absolute to long and back, as previously explained.

Proving termination then becomes difficult, because there is nothing that
precludes a branch instruction from oscillating back and forth between absolute
and long jumps indefinitely.

In order to keep the algorithm in the same complexity class and more easily
prove termination, we decided to explicitly enforce the `branch instructions
must always grow longer' requirement: if a branch instruction is encoded as a
long jump in one iteration, it will also be encoded as a long jump in all the
following iterations. This means that the encoding of any branch instruction
can change at most two times: once from short to absolute (or long), and once
from absolute to long.

There is one complicating factor. Suppose that a branch instruction is encoded
in step $n$ as an absolute jump, but in step $n+1$ it is determined that
(because of changes elsewhere) it can now be encoded as a short jump. Due to
the requirement that the branch instructions must always grow longer, this
means that the branch encoding will be encoded as an absolute jump in step
$n+1$ as well.

This is not necessarily correct. A branch instruction that can be
encoded as a short jump cannot always also be encoded as an absolute jump, as a
short jump can bridge segments, whereas an absolute jump cannot. Therefore,
in this situation we have decided to encode the branch instruction as a long
jump, which is always correct.

The resulting algorithm, while not optimal, is at least as good as the ones
from SDCC and {\tt gcc}, and potentially better. Its complexity remains
the same (there are at most $2n$ iterations, where $n$ is the number of branch
instructions in the program).

\subsection{The algorithm in detail}

The branch displacement algorithm forms part of the translation from
pseudocode to assembler. More specifically, it is used by the function that
translates pseudo-addresses (natural numbers indicating the position of the
instruction in the program) to actual addresses in memory.

Our original intention was to have two different functions, one function
$\mathtt{policy}: \mathbb{N} \rightarrow \{\mathtt{short\_jump},
\mathtt{absolute\_jump}, \mathtt{long\_jump}\}$ to associate jumps to their
intended encoding, and a function $\sigma: \mathbb{N} \rightarrow
\mathtt{Word}$ to associate pseudo-addresses to machine addresses. $\sigma$
would use $\mathtt{policy}$ to determine the size of jump instructions.

This turned out to be suboptimal from the algorithmic point of view and
impossible to prove correct.

From the algorithmic point of view, in order to create the $\mathtt{policy}$
function, we must necessarily have a translation from pseudo-addresses
to machine addresses (i.e. a $\sigma$ function): in order to judge the distance
between a jump and its destination, we must know their memory locations.
Conversely, in order to create the $\sigma$ function, we need to have the
$\mathtt{policy}$ function, otherwise we do not know the sizes of the jump
instructions in the program.

Much the same problem appears when we try to prove the algorithm correct: the
correctness of $\mathtt{policy}$ depends on the correctness of $\sigma$, and
the correctness of $\sigma$ depends on the correctness of $\mathtt{policy}$.

We solved this problem by integrating the $\mathtt{policy}$ and $\sigma$
algorithms. We now have a function
$\sigma: \mathbb{N} \rightarrow \mathtt{Word} \times \mathtt{bool}$ which
associates a pseudo-address to a machine address. The boolean denotes a forced
long jump; as noted in the previous section, if during the fixed point
computation an absolute jump changes to be potentially re-encoded as a short
jump, the result is actually a long jump. It might therefore be the case that
jumps are encoded as long jumps without this actually being necessary, and this
information needs to be passed to the code generating function.

The assembler function encodes the jumps by checking the distance between
source and destination according to $\sigma$, so it could select an absolute
jump in a situation where there should be a long jump. The boolean is there
to prevent this from happening by indicating the locations where a long jump
should be encoded, even if a shorter jump is possible. This has no effect on
correctness, since a long jump is applicable in any situation.

\begin{figure}
\begin{algorithmic}
\Function{f}{$labels$,$old\_sigma$,$instr$,$ppc$,$acc$}
	\State $\langle added, pc, sigma \rangle \gets acc$
	\If {$instr$ is a backward jump to $j$}
		\State $length \gets \mathrm{jump\_size}(pc,sigma_1(labels(j)))$
	\ElsIf {$instr$ is a forward jump to $j$}
		\State $length \gets \mathrm{jump\_size}(pc,old\_sigma_1(labels(j))+added)$
	\Else
		\State $length \gets \mathtt{short\_jump}$
	\EndIf
	\State $old\_length \gets \mathrm{old\_sigma_1}(ppc)$
	\State $new\_length \gets \mathrm{max}(old\_length, length)$
	\State $old\_size \gets \mathrm{old\_sigma_2}(ppc)$
	\State $new\_size \gets \mathrm{instruction\_size}(instr,new\_length)$
	\State $new\_added \gets added+(new\_size-old\_size)$
	\State $new\_sigma_1(ppc+1) \gets pc+new\_size$
	\State $new\_sigma_2(ppc) \gets new\_length$ \\
	\Return $\langle new\_added, pc+new\_size, new\_sigma \rangle$
\EndFunction
\end{algorithmic}
\caption{The heart of the algorithm}
\label{f:jump_expansion_step}
\end{figure}

The algorithm, shown in Figure~\ref{f:jump_expansion_step}, works by folding the
function {\sc f} over the entire program, thus gradually constructing $sigma$.
This constitutes one step in the fixed point calculation; successive steps
repeat the fold until a fixed point is reached.

Parameters of the function {\sc f} are: 
\begin{itemize}
	\item a function $labels$ that associates a label to its pseudo-address;
	\item $old\_sigma$, the $\sigma$ function returned by the previous
		iteration of the fixed point calculation;
	\item $instr$, the instruction currently under consideration;
	\item $ppc$, the pseudo-address of $instr$; 
	\item $acc$, the fold accumulator, which contains $pc$ (the highest memory
		address reached so far), $added$ (the number of bytes added to the program
		size with respect to the previous iteration), and of course $sigma$, the
 		$\sigma$ function under construction.
\end{itemize}

The first two are parameters that remain the same through one iteration, the
final three are standard parameters for a fold function (including $ppc$,
which is simply the number of instructions of the program already processed).

The $\sigma$ functions used by {\sc f} are not of the same type as the final
$\sigma$ function: they are of type
$\sigma: \mathbb{N} \rightarrow \mathbb{N} \times \{\mathtt{short\_jump},
\mathtt{absolute\_jump},\mathtt{long\_jump}\}$; a function that associates a
pseudo-address with a memory address and a jump length. We do this to be able
to ease the comparison of jump lengths between iterations. In the algorithm,
we use the notation $sigma_1(x)$ to denote the memory address corresponding to
$x$, and $sigma_2(x)$ to denote the jump length corresponding to $x$.

Note that the $\sigma$ function used for label lookup varies depending on
whether the label is behind our current position or ahead of it. For
backward branches, where the label is behind our current position, we can use
$sigma$ for lookup, since its memory address is already known. However, for
forward branches, the memory address of the address of the label is not yet
known, so we must use $old\_sigma$.

We cannot use $old\_sigma$ without change: it might be the case that we have
already increased the size of some branch instructions before, making the program
longer and moving every instruction forward. We must compensate for this by
adding the size increase of the program to the label's memory address according
to $old\_sigma$, so that branch instruction spans do not get compromised.

Note also that we add the pc to $sigma$ at location $ppc+1$, whereas we add the
jump length at location $ppc$. We do this so that $sigma(ppc)$ will always
return a pair with the start address of the instruction at $ppc$ and the
length of its branch instruction (if any); the end address of the program can
be found at $sigma(n+1)$, where $n$ is the number of instructions in the
program. 

\section{The proof}

In this section, we present the correctness proof for the algorithm in more
detail.  The main correctness statement is as follows (slightly simplified, here):

\begin{lstlisting}
definition sigma_policy_specification :=
 $\lambda$program: pseudo_assembly_program.
 $\lambda$sigma: Word $\rightarrow$ Word.
 $\lambda$policy: Word $\rightarrow$ bool.
  sigma (zero $\ldots$) = zero $\ldots$ $\wedge$
  $\forall$ppc: Word.$\forall$ppc_ok.
  let $\langle$preamble, instr_list$\rangle$ := program in
  let pc := sigma ppc in
  let instruction :=
   \fst (fetch_pseudo_instruction instr_list ppc ppc_ok) in
  let next_pc := \fst (sigma (add ? ppc (bitvector_of_nat ? 1))) in
   (nat_of_bitvector $\ldots$ ppc $\leq$ |instr_list| $\rightarrow$
    next_pc = add ? pc (bitvector_of_nat $\ldots$
    (instruction_size $\ldots$ sigma policy ppc instruction)))
   $\wedge$     
   ((nat_of_bitvector $\ldots$ ppc < |instr_list| $\rightarrow$
    nat_of_bitvector $\ldots$ pc < nat_of_bitvector $\ldots$ next_pc)
   $\vee$ (nat_of_bitvector $\ldots$ ppc = |instr_list| $\rightarrow$ next_pc = (zero $\ldots$))).
\end{lstlisting}

Informally, this means that when fetching a pseudo-instruction at $ppc$, the
translation by $\sigma$ of $ppc+1$ is the same as $\sigma(ppc)$ plus the size
of the instruction at $ppc$.  That is, an instruction is placed consecutively
after the previous one, and there are no overlaps.

Instructions are also stocked in
order: the memory address of the instruction at $ppc$ should be smaller than
the memory address of the instruction at $ppc+1$. There is one exeception to
this rule: the instruction at the very end of the program, whose successor
address can be zero (this is the case where the program size is exactly equal
to the amount of memory).

Finally, we enforce that the program starts at address 0, i.e. $\sigma(0) = 0$.

Since our computation is a least fixed point computation, we must prove
termination in order to prove correctness: if the algorithm is halted after
a number of steps without reaching a fixed point, the solution is not
guaranteed to be correct. More specifically, branch instructions might be
encoded which do not coincide with the span between their location and their
destination.

Proof of termination rests on the fact that the encoding of branch
instructions can only grow larger, which means that we must reach a fixed point
after at most $2n$ iterations, with $n$ the number of branch instructions in
the program. This worst case is reached if at every iteration, we change the
encoding of exactly one branch instruction; since the encoding of any branch
instructions can change first from short to absolute and then from absolute to
long, there can be at most $2n$ changes.

The proof has been carried out using the ``Russell'' style from~\cite{Sozeau2006}.
We have proven some invariants of the {\sc f} function from the previous
section; these invariants are then used to prove properties that hold for every
iteration of the fixed point computation; and finally, we can prove some
properties of the fixed point.

\subsection{Fold invariants}

These are the invariants that hold during the fold of {\sc f} over the program,
and that will later on be used to prove the properties of the iteration.

Note that during the fixed point computation, the $\sigma$ function is
implemented as a trie for ease of access; computing $\sigma(x)$ is achieved by looking
up the value of $x$ in the trie. Actually, during the fold, the value we
pass along is a pair $\mathbb{N} \times \mathtt{ppc_pc_map}$. The first component
is the number of bytes added to the program so far with respect to
the previous iteration, and the second component, {\tt ppc\_pc\_map}, is a pair
consisting of the current size of the program and our $\sigma$ function.

\begin{lstlisting}
definition out_of_program_none :=
 $\lambda$prefix:list labelled_instruction.$\lambda$sigma:ppc_pc_map.
 $\forall$i.i < 2^16 $\rightarrow$ (i > |prefix| $\leftrightarrow$
  bvt_lookup_opt $\ldots$ (bitvector_of_nat ? i) (\snd sigma) = None ?).
\end{lstlisting}

This invariant states that any pseudo-address not yet examined is not
present in the lookup trie.

\begin{lstlisting}
definition not_jump_default :=
 $\lambda$prefix:list labelled_instruction.$\lambda$sigma:ppc_pc_map.
 $\forall$i.i < |prefix| $\rightarrow$
  ¬is_jump (\snd (nth i ? prefix $\langle$None ?, Comment []$\rangle$)) $\rightarrow$
  \snd (bvt_lookup $\ldots$ (bitvector_of_nat ? i) (\snd sigma)
   $\langle$0,short_jump$\rangle$) = short_jump.
\end{lstlisting}

This invariant states that when we try to look up the jump length of a
pseudo-address where there is no branch instruction, we will get the default
value, a short jump.

\begin{lstlisting}
definition jump_increase :=
 $\lambda$prefix:list labelled_instruction.$\lambda$op:ppc_pc_map.$\lambda$p:ppc_pc_map.
 $\forall$i.i $\leq$ |prefix| $\rightarrow$
 let $\langle$opc,oj$\rangle$ :=
  bvt_lookup $\ldots$ (bitvector_of_nat ? i) (\snd op) $\langle$0,short_jump$\rangle$ in
 let $\langle$pc,j$\rangle$ :=
  bvt_lookup $\ldots$ (bitvector_of_nat ? i) (\snd p) $\langle$0,short_jump$\rangle$ in
  jmpleq oj j.
\end{lstlisting}

This invariant states that between iterations (with $op$ being the previous
iteration, and $p$ the current one), jump lengths either remain equal or
increase. It is needed for proving termination.

\clearpage
\begin{lstlisting}
definition sigma_compact_unsafe :=
 $\lambda$program:list labelled_instruction.$\lambda$labels:label_map.$\lambda$sigma:ppc_pc_map.
 $\forall$n.n < |program| $\rightarrow$
  match bvt_lookup_opt $\ldots$ (bitvector_of_nat ? n) (\snd sigma) with
  [ None $\Rightarrow$ False
  | Some x $\Rightarrow$ let $\langle$pc,j$\rangle$ := x in
    match bvt_lookup_opt $\ldots$ (bitvector_of_nat ? (S n)) (\snd sigma) with
    [ None $\Rightarrow$ False
    | Some x1 $\Rightarrow$ let $\langle$pc1,j1$\rangle$ := x1 in
       pc1 = pc + instruction_size_jmplen j
        (\snd (nth n ? program $\langle$None ?, Comment []$\rangle$)))
    ]
  ].
\end{lstlisting}

This is a temporary formulation of the main property\\
({\tt sigma\_policy\_specification}); its main difference
from the final version is that it uses {\tt instruction\_size\_jmplen} to
compute the instruction size. This function uses $j$ to compute the span
of branch instructions  (i.e. it uses the $\sigma$ function under construction),
instead of looking at the distance between source and destination. This is
because $\sigma$ is still under construction; later on we will prove that after
the final iteration, {\tt sigma\_compact\_unsafe} is equivalent to the main
property.

\begin{lstlisting}
definition sigma_safe :=
 $\lambda$prefix:list labelled_instruction.$\lambda$labels:label_map.$\lambda$added:$\mathbb{N}$.
 $\lambda$old_sigma:ppc_pc_map.$\lambda$sigma:ppc_pc_map.
 $\forall$i.i < |prefix| $\rightarrow$ let $\langle$pc,j$\rangle$ :=
  bvt_lookup $\ldots$ (bitvector_of_nat ? i) (\snd sigma) $\langle$0,short_jump$\rangle$ in
  let pc_plus_jmp_length := bitvector_of_nat ?  (\fst (bvt_lookup $\ldots$
   (bitvector_of_nat ? (S i)) (\snd sigma) $\langle$0,short_jump$\rangle$)) in
  let $\langle$label,instr$\rangle$ := nth i ? prefix $\langle$None ?, Comment [ ]$\rangle$ in
   $\forall$dest.is_jump_to instr dest $\rightarrow$ 
    let paddr := lookup_def $\ldots$ labels dest 0 in
    let addr := bitvector_of_nat ? (if leb i paddr (* forward jump *)
    then \fst (bvt_lookup $\ldots$ (bitvector_of_nat ? paddr) (\snd old_sigma)
     $\langle$0,short_jump$\rangle$) + added
    else \fst (bvt_lookup $\ldots$ (bitvector_of_nat ? paddr) (\snd sigma)
     $\langle$0,short_jump$\rangle$)) in
    match j with
    [ short_jump $\Rightarrow$ $\neg$is_call instr $\wedge$
       \fst (short_jump_cond pc_plus_jmp_length addr) = true
    | absolute_jump $\Rightarrow$ $\neg$is_relative_jump instr $\wedge$
       \fst (absolute_jump_cond pc_plus_jmp_length addr) = true $\wedge$
       \fst (short_jump_cond pc_plus_jmp_length addr) = false
    | long_jump $\Rightarrow$ \fst (short_jump_cond pc_plus_jmp_length addr) = false
       $\wedge$ \fst (absolute_jump_cond pc_plus_jmp_length addr) = false
    ].
\end{lstlisting}

This is a more direct safety property: it states that branch instructions are
encoded properly, and that no wrong branch instructions are chosen.

Note that we compute the distance using the memory address of the instruction
plus its size: this follows the behaviour of the MCS-51 microprocessor, which
increases the program counter directly after fetching, and only then executes
the branch instruction (by changing the program counter again).

\begin{lstlisting}
\fst (bvt_lookup $\ldots$ (bitvector_of_nat ? 0) (\snd policy)
 $\langle$0,short_jump$\rangle$) = 0)
\fst policy = \fst (bvt_lookup $\ldots$
 (bitvector_of_nat ? (|prefix|)) (\snd policy) $\langle$0,short_jump$\rangle$)
\end{lstlisting}

These two properties give the values of $\sigma$ for the start and end of the
program; $\sigma(0) = 0$ and $\sigma(n)$, where $n$ is the number of
instructions up until now, is equal to the maximum memory address so far.

\begin{lstlisting}
(added = 0 $\rightarrow$ policy_pc_equal prefix old_sigma policy))
(policy_jump_equal prefix old_sigma policy $\rightarrow$ added = 0))
\end{lstlisting}

And finally, two properties that deal with what happens when the previous
iteration does not change with respect to the current one. $added$ is a
variable that keeps track of the number of bytes we have added to the program
size by changing the encoding of branch instructions. If $added$ is 0, the program
has not changed and vice versa.

We need to use two different formulations, because the fact that $added$ is 0
does not guarantee that no branch instructions have changed.  For instance,
it is possible that we have replaced a short jump with an absolute jump, which
does not change the size of the branch instruction.

Therefore {\tt policy\_pc\_equal} states that $old\_sigma_1(x) = sigma_1(x)$,
whereas {\tt policy\_jump\_equal} states that $old\_sigma_2(x) = sigma_2(x)$.
This formulation is sufficient to prove termination and compactness.

Proving these invariants is simple, usually by induction on the prefix length.

\subsection{Iteration invariants}

These are invariants that hold after the completion of an iteration. The main
difference between these invariants and the fold invariants is that after the
completion of the fold, we check whether the program size does not supersede
64 Kb, the maximum memory size the MCS-51 can address.

The type of an iteration therefore becomes an option type: {\tt None} in case
the program becomes larger than 64 Kb, or $\mathtt{Some}\ \sigma$
otherwise. We also no longer use a natural number to pass along the number of
bytes added to the program size, but a boolean that indicates whether we have
changed something during the iteration or not. 

If an iteration returns {\tt None}, we have the following invariant:

\clearpage
\begin{lstlisting}
definition nec_plus_ultra :=
 $\lambda$program:list labelled_instruction.$\lambda$p:ppc_pc_map.
 ¬($\forall$i.i < |program| $\rightarrow$
  is_jump (\snd (nth i ? program $\langle$None ?, Comment []$\rangle$)) $\rightarrow$ 
  \snd (bvt_lookup $\ldots$ (bitvector_of_nat 16 i) (\snd p) $\langle$0,short_jump$\rangle$) =
   long_jump).
\end{lstlisting}

This invariant is applied to $old\_sigma$; if our program becomes too large
for memory, the previous iteration cannot have every branch instruction encoded
as a long jump. This is needed later in the proof of termination.

If the iteration returns $\mathtt{Some}\ \sigma$, the invariants
{\tt out\_of\_program\_none},\\
{\tt not\_jump\_default}, {\tt jump\_increase},
and the two invariants that deal with $\sigma(0)$ and $\sigma(n)$ are
retained without change.

Instead of using {\tt sigma\_compact\_unsafe}, we can now use the proper
invariant:

\begin{lstlisting}
definition sigma_compact :=
 $\lambda$program:list labelled_instruction.$\lambda$labels:label_map.$\lambda$sigma:ppc_pc_map.
 $\forall$n.n < |program| $\rightarrow$
  match bvt_lookup_opt $\ldots$ (bitvector_of_nat ? n) (\snd sigma) with
  [ None $\Rightarrow$ False
  | Some x $\Rightarrow$ let $\langle$pc,j$\rangle$ := x in
    match bvt_lookup_opt $\ldots$ (bitvector_of_nat ? (S n)) (\snd sigma) with
    [ None $\Rightarrow$ False
    | Some x1 $\Rightarrow$ let $\langle$pc1,j1$\rangle$ := x1 in
      pc1 = pc + instruction_size
       ($\lambda$id.bitvector_of_nat ? (lookup_def ?? labels id 0))
       ($\lambda$ppc.bitvector_of_nat ?
        (\fst (bvt_lookup $\ldots$ ppc (\snd sigma) $\langle$0,short_jump$\rangle$)))
       ($\lambda$ppc.jmpeqb long_jump (\snd (bvt_lookup $\ldots$ ppc
        (\snd sigma) $\langle$0,short_jump$\rangle$))) (bitvector_of_nat ? n)
       (\snd (nth n ? program $\langle$None ?, Comment []$\rangle$))
    ]
  ].
\end{lstlisting}

This is almost the same invariant as ${\tt sigma\_compact\_unsafe}$, but differs in that it
computes the sizes of branch instructions by looking at the distance between
position and destination using $\sigma$.

In actual use, the invariant is qualified: $\sigma$ is compact if there have
been no changes (i.e. the boolean passed along is {\tt true}). This is to
reflect the fact that we are doing a least fixed point computation: the result
is only correct when we have reached the fixed point.

There is another, trivial, invariant if the iteration returns
$\mathtt{Some}\ \sigma$:

\begin{lstlisting}
\fst p < 2^16
\end{lstlisting}

The invariants that are taken directly from the fold invariants are trivial to
prove.

The proof of {\tt nec\_plus\_ultra} works as follows: if we return {\tt None},
then the program size must be greater than 64 Kb. However, since the
previous iteration did not return {\tt None} (because otherwise we would
terminate immediately), the program size in the previous iteration must have
been smaller than 64 Kb.

Suppose that all the branch instructions in the previous iteration are 
encoded as long jumps. This means that all branch instructions in this
iteration are long jumps as well, and therefore that both iterations are equal
in the encoding of their branch instructions. Per the invariant, this means that
$added = 0$, and therefore that all addresses in both iterations are equal.
But if all addresses are equal, the program sizes must be equal too, which 
means that the program size in the current iteration must be smaller than
64 Kb. This contradicts the earlier hypothesis, hence not all branch
instructions in the previous iteration are encoded as long jumps.

The proof of {\tt sigma\_compact} follows from {\tt sigma\_compact\_unsafe} and
the fact that we have reached a fixed point, i.e. the previous iteration and
the current iteration are the same. This means that the results of
{\tt instruction\_size\_jmplen} and {\tt instruction\_size} are the same.

\subsection{Final properties}

These are the invariants that hold after $2n$ iterations, where $n$ is the
program size (we use the program size for convenience; we could also use the
number of branch instructions, but this is more complex). Here, we only
need {\tt out\_of\_program\_none}, {\tt sigma\_compact} and the fact that
$\sigma(0) = 0$.

Termination can now be proved using the fact that there is a $k \leq 2n$, with
$n$ the length of the program, such that iteration $k$ is equal to iteration
$k+1$. There are two possibilities: either there is a $k < 2n$ such that this
property holds, or every iteration up to $2n$ is different. In the latter case,
since the only changes between the iterations can be from shorter jumps to
longer jumps, in iteration $2n$ every branch instruction must be encoded as
a long jump. In this case, iteration $2n$ is equal to iteration $2n+1$ and the
fixpoint is reached.

\section{Conclusion}

In the previous sections we have discussed the branch displacement optimisation
problem, presented an optimised solution, and discussed the proof of
termination and correctness for this algorithm, as formalised in Matita.

The algorithm we have presented is fast and correct, but not optimal; a true
optimal solution would need techniques like constraint solvers. While outside
the scope of the present research, it would be interesting to see if enough
heuristics could be found to make such a solution practical for implementing
in an existing compiler; this would be especially useful for embedded systems,
where it is important to have as small solution as possible.

In itself the algorithm is already useful, as it results in a smaller solution
than the simple `every branch instruction is long' used up until now---and with
only 64 Kb of memory, every byte counts. It also results in a smaller solution
than the greatest fixed point algorithm that {\tt gcc} uses. It does this
without sacrificing speed or correctness.

This algorithm is part of a greater whole, the CerCo project, which aims to
completely formalise and verify a concrete cost preserving compiler for a large
subset of the C programming language. More information on the formalisation of
the assembler, of which the present work is a part, can be found in a companion
publication~\cite{DC2012}.

\subsection{Related work}

As far as we are aware, this is the first formal discussion of the branch
displacement optimisation algorithm.

The CompCert project is another verified compiler project.
Their backend~\cite{Leroy2009} generates assembly code for (amongst others) subsets of the
PowerPC and x86 (32-bit) architectures. At the assembly code stage, there is
no distinction between the span-dependent jump instructions, so a branch
displacement optimisation algorithm is not needed.

An offshoot of the CompCert project is the CompCertTSO project, who add thread
concurrency and synchronisation to the CompCert compiler~\cite{TSO2011}. This
compiler also generates assembly code and therefore does not include a branch
displacement algorithm.
 
Finally, there is also the Piton stack~\cite{Moore1996}, which not only includes the
formal verification of a compiler, but also of the machine architecture
targeted by that compiler, a bespoke microprocessor called the FM9001.
However, this architecture does not have different
jump sizes (branching is simulated by assigning values to the program counter),
so the branch displacement problem is irrelevant.
  
\subsection{Formal development}

All Matita files related to this development can be found on the CerCo
website, \url{http://cerco.cs.unibo.it}. The specific part that contains the
branch displacement algorithm is in the {\tt ASM} subdirectory, in the files
{\tt PolicyFront.ma}, {\tt PolicyStep.ma} and {\tt Policy.ma}.

\clearpage
\bibliography{biblio}
\bibliographystyle{splncs03}

\end{document}